\documentclass[twocolumn,prl,superscriptaddress]{revtex4}
\usepackage{graphicx}

\def\beq{\begin{equation}}
\def\eeq{\end{equation}}
\newcommand{\ba}{\begin{array}}
\newcommand{\ea}{\end{array}}
\newcommand{\bea}{\begin{eqnarray}}
\newcommand{\eea}{\end{eqnarray} }
\newcommand{\bal}{\begin{align}}
\newcommand{\eal}{\end{align}}
\def\bi{\begin{itemize}}
\def\ei{\end{itemize}}
\def\ben{\begin{enumerate}}
\def\een{\end{enumerate}}
\def\beq{\begin{equation}}
\def\eeq{\end{equation}}
\def\bc{\begin{center}}
\def\ec{\end{center}}
\def\bt{\begin{table}}
\def\et{\end{table}}
\def\btb{\begin{tabular}}
\def\etb{\end{tabular}}
\newcommand{\bvec}{\left ( \ba{c}}
\newcommand{\evec}{\ea \right )}

\def\cl{{\mathcal L}}

\def\mass2{mass${}^2$}

\def\msusy{M_{\rm soft}}

\def\cl{{\mathcal L}}

\def\cH{{\mathcal H}}

\def\mass2{mass${}^2$}

\def\ra{\rangle}
\def\la{\langle}

\def\pa{\partial}

\def\simlt{\stackrel{<}{{}_\sim}}
\def\simgt{\stackrel{>}{{}_\sim}}

\newcommand{\ti}{\tilde}

\def\ov{\overline}

\newcommand{\drawsquare}[2]{\hbox{%
\rule{#2pt}{#1pt}\hskip-#2pt
\rule{#1pt}{#2pt}\hskip-#1pt
\rule[#1pt]{#1pt}{#2pt}}\rule[#1pt]{#2pt}{#2pt}\hskip-#2pt
\rule{#2pt}{#1pt}}

\newcommand{\Yfund}{\drawsquare{7}{0.6}}
\newcommand{\Yasymm}{\drawsquare{7}{0.6}\hskip-7.6pt%
   \raisebox{7pt}{\drawsquare{7}{0.6}}}

\begin{document}

\title{Charming Higgs}%

\author{Brando Bellazzini}
\affiliation{Institute for High Energy Phenomenology,
Newman Laboratory of Elementary Particle Physics,
Cornell University, Ithaca, NY 14853, USA}

\author{Csaba Cs\'aki}
\affiliation{Institute for High Energy Phenomenology,
Newman Laboratory of Elementary Particle Physics,
Cornell University, Ithaca, NY 14853, USA}

\author{Adam Falkowski}
\affiliation{NHETC and Department of Physics and Astronomy, Rutgers University, Piscataway, NJ 08854, USA
}

\author{Andreas Weiler}
\affiliation{CERN, Theory Division, CH-1211 Geneva 23, Switzerland}

\begin{abstract}
We present a simple supersymmetric model where the dominant decay mode of the lightest Higgs boson is $h\to 2 \eta \to 4c$ where $\eta$ is a light pseudoscalar and $c$ is the charm quark.
For such decays the Higgs mass can be smaller than 100 GeV without conflict with experiment.
Together with the fact that both the Higgs and the pseudoscalar $\eta$ are pseudo-Goldstone bosons, this resolves the little hierarchy problem.
\end{abstract}

\maketitle

\def\simgt{\mathrel{\lower2.5pt\vbox{\lineskip=0pt\baselineskip=0pt
          \hbox{$>$}\hbox{$\sim$}}}}
\def\simlt{\mathrel{\lower2.5pt\vbox{\lineskip=0pt\baselineskip=0pt
          \hbox{$<$}\hbox{$\sim$}}}}

\newcommand{\mA}{m_{A^0}}
\newcommand{\Tr}{\rm Tr}
\newcommand{\luv}{\Lambda_{UV}}
\newcommand{\Np}{N^\prime}
\newcommand{\Lh}{\Lambda_H}
\newcommand{\Ls}{\Lambda_S}
\newcommand{\Lg}{\Lambda_G}
\newcommand{\cuFF}{{\cal C}_{m_u}}
\newcommand{\cdFF}{{\cal C}_{m_d}}
\newcommand{\cudF}{{\cal C}_{\mu}}
\newcommand{\cudFb}{\bar{\cal C}_{\mu}}
\newcommand{\cudFF}{{\cal C}_{B_\mu}}

\subsection*{Introduction}

Supersymmetry is probably the most appealing idea for solving the hierarchy problem of particle physics.
Large corrections to the Higgs mass are eliminated due to a symmetry between fermions and bosons which becomes manifest at energies of order the weak scale $v_{EW} = 174$ GeV.
The problem is that many supersymmetric models, in particular the MSSM, imply that some superpartners and the Higgs boson should have already been discovered at LEP.
The absence of such discoveries pushes these models into corners with a relatively large tuning among its basic parameters, of order 1\% or worse.
Naturalness suggests  that supersymmetry, if present at the weak scale, should be combined with additional ingredients.

One possibility is to make the Higgs a pseudo-Goldstone boson (pGB) of a global symmetry broken at the scale $f \simgt v_{EW}$~\cite{otherdouble,double} (see also \cite{bd} for earlier applications of Goldstone bosons in supersymmetric theories).
This idea is usually referred to as {\em double protection} or the {\em super-little} Higgs.
In this class of models, the Higgs and the fermionic sectors are realized in a similar vein as in the little Higgs models \cite{littlehiggs}, in particular collective breaking of the global symmetry is implemented.

Combining supersymmetry with the little Higgs mechanism  further softens quantum corrections to the Higgs mass.
In particular, the one-loop corrections to the Higgs mass are completely finite, making electroweak symmetry breaking technically natural and alleviating the fine-tuning problem of minimal supersymmetry.
The downside of softening quantum corrections are the reduced one-loop contributions to the physical Higgs mass~\cite{Bellazzini:2008zy}.
As a consequence, it is very difficult to make the Higgs mass larger than the LEP bound of 114 GeV without reintroducing fine-tuning or complicating the models by additional structures.
However, we have recently showed~\cite{BCFW} that in this class of models the Higgs mass can be below 114 GeV without conflicting the experimental constraints from LEP.
This is possible because the model predicts non-standard decays for the Higgs.
As advocated for example in~\cite{DG}, the LEP bounds can be relaxed if the Higgs boson decays to a final state with four light standard model (SM) states.

In the model of~\cite{BCFW} the global symmetry breaking pattern is $SU(3)\to SU(2)$ which, in addition to the Higgs doublet,  implies the existence of the fifth pseudo-Goldstone boson $\eta$ who is a singlet under the SM gauge symmetries.
This $\eta$ is naturally light, lighter than the Higgs boson, and has tree-level  derivative couplings to the Higgs suppressed by $v_{EW}/f$-suppressed.
When $f$ is not much larger than the electroweak scale then the decay $h \to \eta \eta$ dominates, while the branching ratio for the standard $h \to b \bar b$ channel can be suppressed below 20\% which is consistent with LEP bounds for a Higgs mass of order $m_Z$.
If this is the case then the dominant Higgs decay channel involves at least four SM states.
The composition of the final state crucially depends on the mass and couplings of the intermediate state $\eta$.
If $\eta$ is heavier than twice the bottom quark mass then Higgs decays into a 4b final state which is excluded by LEP for Higgs masses below 110 GeV \cite{LEP4b}.
But when $\eta$ is lighter than $\sim 2 m_b$  it decays into 2 gluons via a loop of off-shell bottom quarks.
Then the dominant Higgs decay channel is  $h\to 2\eta \to 4 g$, which allows us to evade all existing LEP bounds with the Higgs as light as 80-90 GeV \cite{OPAL4g}.
Incidentally, such a 4 gluon jet final state is difficult to see not only at LEP but also at hadron colliders such as Tevatron and the LHC where it is {\em buried} under the overwhelming QCD background.

In this paper we consider an alternative implementation of the doubly protected Higgs which leads to a different phenomenology.
The present model has the same gauge symmetry $SU(3)_C\times SU(3)_W \times U(1)_X$ and the global symmetry breaking $SU(3)\to SU(2)$ that leads to 5 pseudo-Goldstone bosons: the Higgs doublet and a light pseudoscalar $\eta$.
The main new feature is the embedding of the SM fermions into representations of the gauge group; roughly, the representations of the up- and down-type quarks and leptons are interchanged with respect to \cite{BCFW}.
The striking consequence of that assignment is that the couplings of $\eta$ to the down type quarks and charged leptons can be very suppressed.
If that is the case, $\eta$ does not decay to bottom quarks or tau leptons even if it is kinematically allowed.
Thus, unlike in all previous models of the light hidden Higgs, the pseudoscalar mass does not have to be squeezed into a small window ({\it few} GeV $ <  m_\eta < 2 m_b$) in order to avoid the stringent LEP bounds on the 4b final state;  instead, all the parameter space up to half the Higgs mass is available.

As the decay modes $\eta \to b\bar{b}$ and  $\eta \to \tau \bar \tau$ are suppressed, the branching ratio for $\eta$ decaying into two {\em charm quarks} is by far the largest.
The dominant Higgs decay channel is then $h\to 2\eta \to 4c$ for which the LEP bounds are very similar as for the 4g final state \cite{OPAL4g}.
The branching ratio for $h\to 2\eta \to 2c 2g$ (where $\eta$ decay to gluons now proceeds mainly via a loop of charm quarks and its symmetry partners) is at the level $10^{-2}-10^{-1}$, while the branching ratio for decays with two photons in the final state is even more suppressed, at the level of $10^{-5}-10^{-4}$.
Since charm tagging is difficult in hadron colliders such as the LHC and the Tevatron, the {\em charming Higgs} may well be {\em buried} under the QCD background unless dedicated search strategies are devised.

\subsection*{Gauge sector, symmetry breaking and Goldstone bosons}

We consider a supersymmetric model with the Higgs arising as a pseudo-Goldstone boson of an approximate $SU(3)$ global symmetry spontaneously broken to $SU(2)$.
The global $SU(3)$ is a residue of an extended gauge symmetry broken at higher energies of order 10 TeV.
In our model, the SM gauge symmetry is extended to $SU(3)_C \times SU(3)_W \times U(1)_X$ which is then broken by two pairs of triplet Higgses $\cH_u\,,\,\Phi_{u} =(1,\bar{3})_{1/3}$ and $\cH_d\,,\,\Phi_{d} =(1,3)_{-1/3}$ with the following charges:
\beq
\label{e.pgb}
\begin{array}{c|c|c|c|}
& SU(3)_C & SU(3)_W & U(1)_X
\\
\hline
\cH_u,\Phi_u & 1 & \bar 3 &  1/3
\\
\cH_d,\Phi_d  & 1 &  3 & -1/3
\end{array}\,.
\eeq
We assume that the $\Phi$'s and $\cH$'s do not mix in the superpotential.
This leads to an enlarged SU(3)$_\Phi$ $\times$SU(3)$_\cH$  approximate global symmetry where the two group factors independently rotate the respective  triplet pair. The $\Phi$'s are assumed to have a supersymmetric VEV:
\beq
\la \Phi_u \ra =  \la \Phi_d \ra^T  =(0,0,F/\sqrt{2})
\eeq
with $F \sim 10$ TeV.
This breaks the gauge group down to $SU(3)_C \times SU(2)_W \times U(1)_Y$ with the hypercharge realized as $Y = - T_8/\sqrt{3} + X$.
On the other hand, SU(3)$_\cH$ survives down to lower energies.
Ultimately, loops involving the top quark and its symmetry partners generate a negative mass squared for $\cH_{u,d}$ (and also the quartic term) which induces a VEV of $\cH_{u,d}$ of order $M_{\rm soft}$.
Then the approximate global SU(3)$_{\cH}$ symmetry is spontaneously broken to SU(2) and produces 5 pseudo-Goldstone bosons (pGBs).
Four of them corresponds to the Higgs doublet whose 3 components are non-physical and eaten by the W and Z bosons.
This leaves two physical pGBs.
It is convenient to use the following embedding of these two pGBs into the Higgs triplets:
\begin{eqnarray}
\label{e.htp}
\cH_u^T &=& f s_b \bvec \sin(\ti h/\sqrt {2}f) \\ 0 \\ e^{i \ti \eta/\sqrt {2}f} \cos(\ti h/\sqrt {2}f) \evec , \nonumber \\
\cH_d &=& f c_b  \bvec \sin(\ti h/\sqrt {2}f) \\ 0 \\ e^{-i \ti \eta/\sqrt {2}f} \cos(\ti h/\sqrt {2}f) \evec\,.
\end{eqnarray}
In the above $c_b = \sqrt{1-s_b^2}$ and $t_b \equiv s_b/c_b$ is the analogue of the MSSM $\tan \beta$.
The field $\tilde{h}$ is the pGB Higgs whose VEV will break the electroweak symmetry.
The other physical pGB  $\ti \eta$ is a singlet under the SM gauge interactions.
The Higgs boson field $h$ is obtained by the shift $\tilde{h}=h+\sqrt{2} \tilde{v}$,
while the canonically normalized singlet is  $\eta = \ti \eta \cos(\ti v/f)$.
Once the fermions are introduced (see the next section) the non-linear sigma model scale $f$ is generated dynamically by loops of the top quark and its symmetry partners, in close analogy to generation of the Higgs vev in the MSSM.
We are interested in the case where $f$ is not too large, of order $350-400$ GeV which requires some mild tuning among the model parameters.
The radial mode corresponding to the oscillations of $f$ (which is not a pGB) has a mass of order 200-300 GeV.
The top/stop loops also generate the VEV $\la \ti h \ra  = \sqrt{2} \ti v$  of the Higgs field.
The electroweak scale is related to the Higgs VEV  by
\beq
v_{EW} = f \sin(\ti v/f), 
\eeq
and the Higgs mass ends up in the range 80-90 GeV for a generic point in the parameter space.

\subsection*{Matter fields}

The third generation quarks and leptons are embedded into the following anomaly free representations:
\begin{equation}
\begin{array}{l|c|c|c|}
& SU(3)_C &  SU(3)_W & U(1)_X
\\
\hline
Q = (t^Q,b^Q,\hat t^Q)& 3 & 3 & 1/3
\\
t_c^{1,2} & \bar 3 & 1 & -2/3\\
b_{c} & \bar 3 & 1 & 1/3\\ \hline
L_{1,2} = (\tau_{1,2}, \nu_{1,2},\hat \tau_{1,2}) &  1 & \bar 3 & -2/3 \\
L_c =   (\nu_c^L,\tau_c^L,\hat \nu_c^L )&  1 & \bar 3 & 1/3 \\
\tau_{c}^{1,2,3} &  1 & 1 & 1
\end{array}
\end{equation}
The third generation quark sector is fairly simple, in fact it coincides with the extended quark sector of common little Higgs models. Compared to the SM, only one heavy vector-like top quark is added. This matter content can be obtained from an underlying $SU(6)\times U(1)_X$ symmetry, see the Appendix.
The masses for all quarks can be obtained from the superpotential
\beq
\label{e.yc}
y_1 t_c^1 \Phi_u  Q  +  y_2 t_c^2 \cH_u  Q   +  {y_{b} \over \mu_V}  b_c^1 Q \Phi_d \cH_d.
\eeq
The SM top mass follows
\begin{equation}
m_t \approx \frac{s_b y_1 y_2 F}{\sqrt{(y_1 F)^2+ 2 (s_b y_2 f)^2}} v_{EW},
\end{equation}
while the mass of the heavy top partner is
\beq
m_T = \sqrt{(y_1 F)^2/2 + (s_b y_2 f)^2}.
\eeq
Note that the couplings in Eq. (\ref{e.yc}) are non-generic, as the gauge symmetry also allows  for
\beq
\label{e.ync} 
\ti y_1 t_c^1 \Phi_u  Q  +  \ti y_2 t_c^2 \cH_u  Q.
\eeq
We assume that the tilded Yukawas are absent or small (which in practice means $10^{-2}$ or less).
In this case the approximate  global $SU(3)_{\cH}$ symmetry acting on  $\cH_{u,d}$ is only {\em collectively} broken by the renormalizable couplings.
That is to say, SU(3)$_{\cH}$ would be unbroken by (\ref{e.yc}) if $y_2 = 0$, and also if $y_1 = 0$ (in which case the global rotation of $\cH_u$ could be compensated by a global rotation of $Q$).
Thus, the order parameter for the explicit breaking of the global $SU(3)_\cH$ symmetry is proportional to ${\rm max}(y_1 F,y_2 f)$.
When the collective breaking is implemented the top loop contributions to the pGB Higgs mass are finite at one loop in a supersymmetric theory because they have to be proportional to $\msusy^2$ and there is simply no room for another mass parameter.
In order to keep the size of loop corrections under control both $y_2 f$ and $y_1 F$ have to be below 1 TeV (since $F\sim 10$ TeV the latter  requires $y_1 \simlt 0.1$).
Given these assumptions, electroweak symmetry breaking has no fine-tuning at all.

The present choice of the quark content is somewhat more appealing than that in \cite{BCFW}.
While the bottom quark mass originates from non-renormalizable couplings, this can be easily cured by integrating in a vector-like pair of $SU(3)_W$ and color triplets $V$, $V_c$ with zero $U(1)_X$ charge, and adding the couplings $\mu_V V_c V  + y_{b1}  V_c Q \Phi_d +  y_{b2}  b_c V \cH_d$ to the superpotential.
For large $\mu_V$ this leads to the effective non-renormalizable interaction in (\ref{e.yc}) with $y_b = y_{b1} y_{b2}$ and the resulting bottom mass is $m_b \approx y_b c_b v_{EW} F/\sqrt 2 \mu_V$.
Due to the fact that the bottom quark is much lighter than the electroweak scale we are free to choose large $\mu_V$, as large as $10-100$ TeV. 
Without including $V,V_c$ the 1-loop $\beta$-function for QCD vanishes, while with these fields added it is positive which leads to a Landau pole below the GUT scale. 
Typically, there is also a Landau pole for the $y_2$ Yukawa coupling which should be large in order to minimize fine tuning. 
The location of the Landau pole $\Lambda = {\rm min}[\Lambda_{QCD},\Lambda_{y_2}]$ varies between $10^3$ and $10^6$ TeV.

For the second generation quarks 
we assume that the same matter structure is repeated, just replacing the Yukawa couplings $y_i \to y_{ci}$.
In particular, the charm mass is given by
\beq
m_c =  {s_b y_{c1}  y_{c2} F \over \sqrt{ (y_{c1} F)^2 + 2 (s_b y_{c2} f)^2 }}  v_{EW}\,.
\eeq
One can see that there are two ways to make the charm quark mass hierarchically smaller than $v_{EW}$: either $y_{c1} F \ll y_{c2} f $  (where $y_{c2} f > 400$ GeV to make the charm partner heavy enough) or the other way around.
In the following we choose the former possibility; the latter leads to very suppressed couplings of $\eta$ to the charm quarks and similar phenomenology as in \cite{BCFW}.

The charged lepton sector is somewhat more complicated, which is the price we have to pay for canceling the anomalies.
In particular, the SM tau lepton has three heavy partner states, while in the neutrino sector we have one heavy partner state; note that Majorana masses are not allowed by this representation assignment.
In the lepton sector we have multiple options for writing the Yukawa couplings consistent with the collective symmetry breaking.
Here  we concentrate on the following set of Yukawas:
\begin{equation}
\label{e.tauycmod} 
\alpha_{1j} \tau_c^j L_1 \Phi_d  + \alpha_{2j} \tau_c^j L_2 \Phi_d
+ \beta_2 \Phi_u  L_c L_2 + \tilde \alpha_{13} \tau_c^3 L_1 \cH_d \ .
\end{equation}
This is a collective Yukawa, since a global SU(3) emerges if one sets either $\tilde{\alpha}_{13}$ or $\alpha_{1j}$ to zero.  In the absence of the $\tilde{\alpha}_{13}$ term all $\tau$-partners pick up a mass proportional to $F$ and are pushed into the TeV range.
Including  $\tilde{\alpha}_{13}$ provides the mass for the SM $\tau$.
At this point the neutrino mass matrix has one zero eigenvalue.
This can be taken care of by adding the coupling $\beta_1 \cH_u L_c L_1$ with $\beta_1 \sim 10^{-13}$.

\subsection*{Higgs decays: $h\to 2 \eta$ vs. $h\to b\bar{b}$}

We first discuss the Higgs decay modes, and argue that the $h\to \eta \eta$ mode dominates if the global symmetry scale $f$ is not much larger than the electroweak scale.
Even though  $\eta$ is a complete SU(2) singlet, it does have a tree-level derivative coupling to the Higgs field $h$ due to $h$ partly living in the third component of $\mathcal{H}_{u,d}$. The symmetry preserving derivative coupling (characteristic to exact Goldstone bosons) originates from the Higgs kinetic terms,
\beq
\cl_{pGB} \approx {1 \over 2}(\pa_\mu \ti h)^2 + {1 \over 2} \cos^2(\ti h/\sqrt{2} f) (\pa_\mu \ti \eta)^2.
\eeq
After canonical normalization of the pGB fields this leads to the following vertex of the Higgs boson with two singlets:
\beq
\cl_{h\eta^2} \approx - h (\pa_\mu \eta)^2  {\tan(\ti v/f) \over \sqrt{2} f}.
\eeq
The decay width of the Higgs boson into two singlets is given by
\beq
\Gamma_{h \to \eta\eta} \approx {1 \over 64 \pi} \sqrt{1 - {4 m_\eta^2 \over m_g^2}}\left( 1 - {v_{EW}^2 \over f^2} \right)^{-1}
{m_h^3 v_{EW}^2\over f^4}.
\eeq
The coupling of the Higgs boson  to the SM quarks and leptons is the same as in  the SM, up to an additional factor $\cos (\ti v/f)$ that arises due to its pGB nature (c.f. the sines and cosines in the parametrization Eq.(\ref{e.htp})).
Thus, the decay width into a pair of SM fermions is given by
\beq
\Gamma_{h \to f \ov f}
=  \cos^2(\ti v/f)\Gamma_{h \to f \ov f}^{SM}
= c_{QCD} {N_c^f \over 16 \pi}  \cos^2(\ti v/f) {m_h m_f^2 \over v_{EW}^2},
\eeq
where  $N_c^f = 3$ for quarks and $1$ for leptons.
The factor $c_{QCD}$ arises due to  higher order QCD corrections which  can be numerically important; for example for the b-quark it is given by $c_{QCD} \approx 1/2$ for a 100 GeV Higgs \cite{DKS}.
The relevant quantity for LEP searches, customarily denoted as $\xi^2_{h\to b \ov b}$,  is the branching ratio for a decay into $b$ quarks multiplied by the suppression of the Higgs production cross section.
The latter is also relevant here because the coupling of the pGB Higgs to the Z boson is multiplied, much as the Higgs-fermion coupling, by the factor  $\cos(\ti v/f) < 1$ as compared to the SM coupling.
It then follows
\beq
\xi^2_{h \to b \ov b}
= {\Gamma^{SM}_{h \to b \ov b} \over \Gamma_{h \to \eta\eta}
+ \left(1 - {v_{EW}^2 \over f^2}\right) \sum_f \Gamma^{SM}_{h \to f \ov f}}
\left(1 - {v_{EW}^2 \over f^2}\right)^2.
\eeq
If $f$ is as small as 350-400 GeV, the $b \ov b$ branching ratio is at the level of 10-20 percent for the Higgs mass of order $m_Z$.
That is small enough to avoid exclusion by LEP.
Once $f$ is raised to around $450$ GeV or higher, the generic 114.4 GeV limit from LEP cannot be significantly relaxed - the $b \ov b$ branching ratio becomes large enough to have been observable at LEP.

\subsection*{$\eta$ couplings and decays}

The pGB pseudoscalar $\eta$ has linear couplings to the SM fermions of the form 
\beq
\label{e.ec}
i {\ti y_f} \eta \ov f  \gamma_5 f \,.
\eeq
The partial width for $\eta \to f\bar{f}$ is given by
\beq
\Gamma(\eta\to f \bar f)= N_c^f \sqrt{1 - {4 m_f^2 \over m_\eta^2} } {m_\eta \over 8 \pi} \ti y_f^2.
\eeq
At tree-level, the bottom quark does not couple to $\eta$ at all in the effective theory below $\mu_V$, c.f. Eq. (\ref{e.yc}) (a tiny coupling suppressed by $\mu_V$ is generated when we integrate in the vector-like pair $V,V_c$).
The leading coupling of the form (\ref{e.ec}) is generated by a penguin diagram involving two top quarks and the W boson.
One can estimate
\beq
\ti y_b \sim {1 \over 16 \pi^2} \ti y_t {m_t m_b \over v_{EW}^2} \log(m_T^2/m_t^2) \sim 10^{-4},
\eeq
where $\ti y_t \sim 0.2$ is the coupling of $\eta$ to the top quark.
The loop factor provides enough suppression of $\ti y_b$ so that $\eta$ does not decay into two bottom quarks even when it is kinematically allowed!
This interesting feature distinguishes our set-up from all previous hidden Higgs models in the literature where decays to b quarks could be avoided only for $m_\eta < 2 m_b$.

For the charm quarks, the coupling to $\eta$ originates from  the term $s_b y_{2c} f \cos (\ti v/f) e^{i\ti \eta /\sqrt 2 f}  (c_c^2 \hat c^Q)$, while for the tau lepton the relevant term is $c_b \ti \alpha_{13} f \cos (\ti v/f) e^{-i\ti \eta /\sqrt 2 f}  (\tau_c^3 \hat \tau_1)$.
Expressing the original fields in terms of mass eigenstates one finds
\beq
\ti  y_c \approx {m_c \over \sqrt 2 f},
\qquad
\ti y_\tau  \sim {m_\tau \over \sqrt 2 f} {m_\tau^2 \over M_\tau^2}.
\eeq
For $f \sim 350$ GeV we find $\ti y_c \sim 10^{-3}$, while $\ti y_\tau$ is additionally suppressed by the ratio of the tau mass to its heavy partner mass, $(m_\tau/M_\tau)^2 \sim 10^{-5}$.
Given the loop suppression of  $\ti y_b$, the charm coupling $\ti y_c$ remains by far the largest coupling.
Therefore the dominant decay channel of the pseudoscalar is $\eta \to c \bar c$, with the total width of order  keV.

\begin{figure}
\includegraphics[scale=0.8]{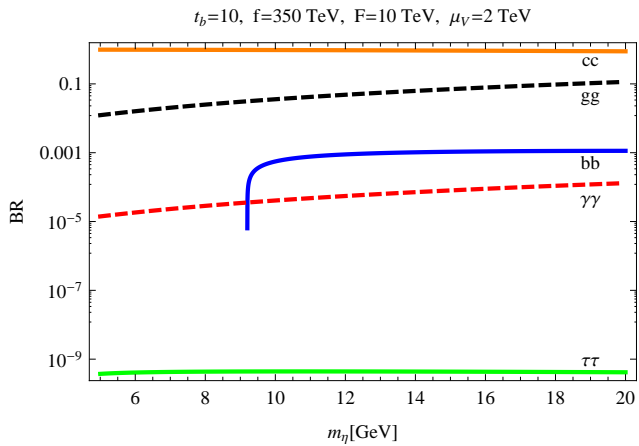}
\caption{The branching ratios of $\eta$ into bottom (blue), tau (green), charm (orange), gluons (dashed black) and photons (dashed red) as a function of its mass.
We picked the following generic point in Yukawa coupling space:
$y_1 = 0.109$, $y_2 = 1.8$,  $y_{b1} = 0.3$, $y_{b2} = 0.365$, $y_{c1} = 0.0003$, $y_{c2} = 0.1$, $\alpha_{11}= \alpha_{22} = \beta_2  = 0.1$, $\ti \alpha_{13} = 0.102$, while the remaining Yukawas are set to zero.
\label{fig:BRs}}
\end{figure}

In addition to tree-level decays, the $\eta$ can decay to two gluons or photons via triangle diagrams with light or heavy fermions in the loop.
In Fig.~\ref{fig:BRs} we present the branching ratios of $\eta$ for a typical point in the Yukawa space. In the entire range of $\eta$ masses, the decay into charm quarks dominates, with the next-to-leading decay into two gluons suppressed by a factor of 100. The decay widths into photons is suppressed at the $10^{-4}-10^{-5}$ level, while decays to bottom quarks (even when  kinematically allowed) and to $\tau$'s are extremely suppressed.

\subsection*{Conclusions}

We presented a supersymmetric model where the lightest Higgs boson decays dominantly into four charm quarks.
This decay is mediated by two on-shell pseudoscalars $\eta$, each subsequently decaying into two charm quarks.
Besides the interesting phenomenology, our model is motivated by solving the fine-tuning problem of minimal supersymmetric theories.
The Higgs is a pGB of a spontaneously broken global symmetry, and it is protected against divergent quantum corrections at one loop, including logarithmic divergences.
This opens up the possibility of  completely natural electroweak symmetry breaking.
The softening of the quantum correction implies that the Higgs boson cannot be much heavier than 80-90 GeV without reintroducing fine-tuning.
This mass range is however perfectly allowed by all existing constraints thanks to the fact that the  $h \to 4 c$ decay channel is poorly constrained by the existing LEP analyses.

In our model, decays of $\eta$ to bottom quarks are significantly suppressed (even when they are kinematically allowed), while decays into $\tau$'s are extremely suppressed in most of the parameter space.
To our knowledge, this peculiar pattern of the Higgs branching ratios is not available in any other model in the literature.

Signatures of Higgs cascade decays are currently searched for at the Tevatron \cite{tev} in the 4$\tau$ and $2 \tau 2 \mu$ channels.
These channels are motivated by the hidden Higgs models based on the NMSSM \cite{DG}.
The existence of well motivated hidden Higgs models with suppressed decays to bottom quarks and tau leptons prompts extending the scope of collider searches.
The feasibility of detecting fully hadronic final states like $4c$ and $4g$ should be assessed.
Moreover, the searches should cover a larger range of the intermediate pseudoscalar masses, including $m_\eta > 2 m_b$.
The flip side of the coin is that $\eta$ heavier than $2 m_b$ cannot be probed at B-factories, unlike in the previous hidden Higgs models \cite{BaBar}.

\begin{figure}
\includegraphics[scale=0.6]{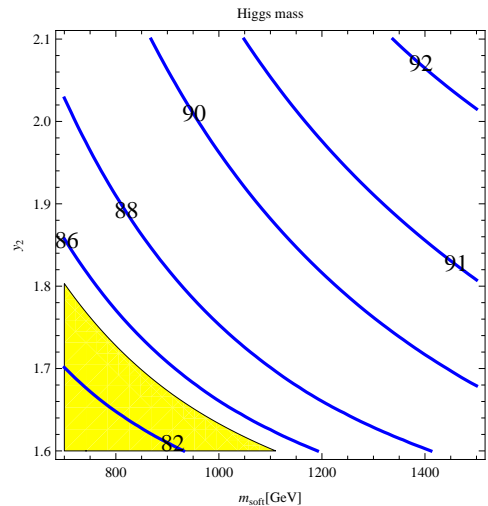}
\caption{The contours of the Higgs mass as a function of the top Yukawa coupling $y_2$ and the soft mass scale $M_{\rm soft}$. The yellow region is excluded by $h \to b \ov b$ searches at LEP. The Landau pole is at around $10^6$ TeV for $y_2\sim 1.8$, while it is lowered to around $10^3$ TeV for $y_2 \sim 2.1$.
\label{fig:widths}}
\end{figure}

\section*{Acknowledgements}
We would like to thank Tao Han and Itay Yavin for useful discussions.
The work of A.F. was supported in part by the Department of Energy grant DE-FG02-96ER40949. This research of B.B. and C.C. has been supported in part by the NSF grant PHY-0757868. C.C. was also supported in part by a U.S.-Israeli BSF grant.

\subsection*{Appendix: The flipped matter content}

It is well-known that an alternative choice for embedding the SM matter content into an SU(5) group is to use the {\em flipped} SU(5)$\times$U(1)$_X$ gauge group with representation $10_1+\bar{5}_{-3}+1_5$, where a right-handed neutrino is included, and hypercharge is obtained as $Y=\frac{1}{30} {\rm diag}(2,2,2,-3,-3)+ \frac{X}{5}$. The main effect is to flip the role of up- and down-type quarks, for example $\bar{5}\to \bar{d}_R + L_L$, hence the name flipped SU(5). Interestingly, this flipped matter content can be generalized to any SU(N) group with the following anomaly free representation containing the chiral matter content of one family (the first number indicates the multiplicity of each representation):
\begin{equation}
\begin{array}{ccc} & SU(N) & U(1)_X \\ \hline
1 & \Yasymm & N-4 \\
(N-4)& \overline{\Yfund} & -(N-2) \\
\frac{(N-4)(N-3)}{2} & 1 & N \end{array}
\end{equation}
The anomaly cancellation of this generalized flipped matter does not seem to be readily following from the anomalies of an SO(2N) spinor (as it does in the case of SU(5)). The particular case of SU(6)$\times$U(1)$_X$ (with matter $\Yasymm_1+2\times \overline{\Yfund}_{-2}+3\times 1_3$) contains the SU(3)$_{QCD}\times$SU(3)$_W\times$U(1)$_Z\times$U(1)$_X$ subgroup, where hypercharge is identified as $Y=X/3-T_8/6$ with $T_8={\rm diag} ( 1,1,-2)$ of SU(3)$_W$. This is the way the chiral part of the matter content used in this paper has been obtained.

\end{document}